\documentclass{ws-ijgmmp}
\usepackage{cite}
\begin{document}

\markboth{S. Christina, T. Ibungochouba Singh and I. Ablu Meitei}
{Modified Hawking temperatures of black holes in Lorentz violation theory}

%
\catchline{}{}{}{}{}
%

\title{MODIFIED HAWKING TEMPERATURES OF BLACK HOLES IN LORENTZ VIOLATION THEORY}

\author{S. CHRISTINA}

\address{Department of Mathematics, Manipur University, Canchipur\\ Imphal West, Manipur, India}

\author{T. IBUNGOCHOUBA SINGH\footnote{T. IBUNGOCHOUBA SINGH
}}

\address{Department of Mathematics, Manipur University, Canchipur\\
Imphal West, Manipur, India\,\footnote{Department of Mathematics, Manipur University, Canchipur
Imphal West, Manipur, India}\\
\email{ibungochouba@rediffmail.com\footnote{ibungochouba@rediffmail.com}} }

\author{I. ABLU MEITEI}

\address{Department of Physics, DM College of Science,  Manipur, India\\
}
\maketitle

\begin{history}
\received{(Day Month Year)}
\revised{(Day Month Year)}
\end{history}

\begin{abstract}
In this paper, the tunneling of scalar particles near the event horizons of Riemann space time, BTZ black hole and Schwarzschild-de Sitter black hole are investigated by applying Hamilton-Jacobi equation with Lorentz violation theory in curved space time. The modified Hamilton-Jacobi equation is derived from Klein-Gordon equation of scalar particles induced by Lorentz violation theory in curved space time. The Hawking temperatures of Riemann space time and the BTZ black hole are modified due to the effect of Lorentz violation theory. Moreover, the Bekenstein-Hawking entropy near the event horizon of Schwarzschild-de Sitter black hole is also modified due to Lorentz violation theory. It is observed that the modified values of Hawking temperatures and change in Bekenstein-Hawking entropy depend upon the ether-like vectors $u^\alpha$.
\end{abstract}

\keywords{Hawking radiation; Klein-Gordon equation; Bekenstein-Hawking entropy; Lorentz violation theory in curved space time; ether-like vector.}

\section{Introduction}	
As per the principles of classical theory, black holes do not emit particles. But in 1974, there was a paradigm shift in black hole physics when Stephen Hawking discovered that black holes can emit particles using the techniques of quantum field theory in curved space time \cite{haw1, haw2}. Since then, many researches have paid more attentions to the study of Hawking radiation of black holes. Bekenstein \cite{bek1, bek2} and Bardeen et. al \cite{bar} developed the black hole thermodynamics and showed that the black hole entropy is proportional to the area of the event horizon. Damour and Ruffini \cite{dam} proposed a method for studying Hawking radiation using tortoise coordinate transformation where the gravitational field is taken to be independent of time. The Dirac equations and Maxwell's electromagnetic field equations expressed in Newman-Penrose formalism and Klein-Gordon equation are transformed to a standard wave equation at the event horizon by using tortoise co-ordinate transformation and Newman-Penrose spin coefficients. Separating the wave equation, the incoming and the outgoing wave equations are derived and corresponding surface gravity and Hawking temperatures at the event horizon of black holes are obtained. Following this method, many authors have obtained interesting results in \cite{wu1,wu2,ibo, lanj, ibuab, ibung}.

Further, various methods have been proposed for the study of Hawking radiation as a tunneling process caused by quantum fluctuations across the event horizon of the black hole. According to Kraus and Wilczek \cite{kra1, kra2} and Parikh and Wilczek \cite{par}, the potential barrier is created by the outgoing particle and subsequently the imaginary part of the radial action is derived in accordance with the semiclassical approximation. Zhang and Zhao \cite{zha1,zha2,zha3} further extended this method to a charged rotating black hole and showed that the spectrum is no longer purely thermal. Angheben et. al \cite{ang} put forward a slightly different method as an extension of Padmanbhan's works \cite{sri}. In this method, Hawking radiation is investigated as a tunneling for the extremal rotating black hole using the relativistic Hamilton-Jacobi equation, WKB approximation and Feynman prescription without considering the back reaction of the emitted particles. The Hawking radiation in different types of black holes using Hamilton-Jacobi equation have also been discussed in \cite{yan,wan,chen,ibu1}. Kerner and Mann \cite{ker} investigated the Hawking radiation of black hole by applying Dirac equation, Pauli Sigma matrices, Feynmann prescription and  WKB approximation. Choosing appropriate Gamma matrices from the line element and inserting suitable wave function into the Dirac equation, the action of the radiant particle is obtained. Refs. \cite{ban1, ban2} discussed the Hawking temperature as a tunneling by applying the Hamilton-Jacobi equation beyond the semiclassical approximation. Using first law of black hole mechanics, correction to Benkenstein-Hawking area law containing logarithmic term is obtained. Applying their method, many interesting results have been derived in \cite{maj,ibu2,akh2,dou}. Kruglov \cite{kru1, kru2} investigated the tunneling of vector boson particles by applying WKB approximation, Feynmann prescription and Hamilton-Jacobi ansatz to Proca equation. Using Proca equation, the tunneling of vector boson particles in complicated black holes have also been discussed in \cite{ger,sak1,sak2,ibu3,gonz,ovg1}. {\bf Ref. \cite{ovg2} investigated the Hawking temperatures for different black holes by applying the topological method. Akhmedova et. al \cite{akh1} investigated Hawking radiation using quasi-classical calculation and showed that quasi-classical calculations are associated with temporal contribution and no analogue for the temporal contribution exists in the quantum mechanical tunneling problem.}
Quantum gravity theories such as string theory, loop quantum gravity theory and Gedankan experiments \cite{tow, ven} indicate the existence of a minimum observable length in Planck scale which leads to the generalised uncertainty principle (GUP). Based on GUP, the tunneling of scalar particles and fermions near the event horizon of black holes have been studied in \cite{ablu,ibu4,kene, waji}. It is believed that Lorentz symmetry, which is the foundation of general relativity, may break at high energy level. Therefore, many researchers \cite{liu1,zha4,liu2,liu3} have proposed gravity models on the basis of Lorentz symmetry violation theory. Refs. \cite{cas, nas} have introduced ether-like vectors $u^{\alpha}$ for the study of Lorentz symmetry violation of Dirac equation in flat space time. Refs. \cite{hor,lin,kos1,kos2,blu,jac,onik} have studied the modified Hawking radiation and entropy of black holes by using Hamilton-Jacobi equation with Lorentz violation theory in curved space time.

The objective of the paper is to study the modified Hawking temperatures of scalar particles crossing the event horizons of Riemann space time and BTZ black hole by considering the Lorentz violation theory in curved space time. The paper also intends to study the modified Bekenstein-Hawking entropy of the Schwarzschild-de Sitter black hole. For this purpose, we utilise the modified Hamilton-Jacobi equation in Lorentz violation theory.

This paper is organised as follows. In Section 2, the modified Hamilton-Jacobi equation is revisited in Lorentz violation theory. In Section 3, Hawking temperature of Riemann space time is modified by using modified Hamilton-Jacobi equation induced by Lorentz violation theory in curved space time. The modified Hawking temperatures for uncharged and charged non-rotating BTZ black hole are studied by using modified Hamilton-Jacobi equation in Section 4 and Section 5 respectively. In Section 6, the change in Bekenstein-Hawking entropy is studied by using the modified Hamilton-Jacobi equation. In Section 7, discussion and conclusion are given.

\section{Revisiting the modified Hamilton-Jacobi equation }

Based on high energy quantum gravity theory, Lorentz invariance needs to be corrected at the Planck scale. According to Ref. \cite{gom}, the Lagrangian $\mathcal{L}$ in the new Lorentz violation equation of scalar particles with mass $m$ in the presence of ether-like vectors $u^\alpha$ is
\begin{eqnarray}
\mathcal{L}=\frac{1}{2}[\partial_\mu \psi \partial^\mu\psi+\lambda(u^\alpha\partial_\alpha \psi)^2+m^2\psi^2],
\end{eqnarray}
where $\lambda$ is a proportionality constant and the ether like vectors $u^\alpha$ satisfy
\begin{eqnarray}
u^\alpha u_\alpha=const.
\end{eqnarray}
Therefore, in accordance with the principle of least action, the corrected scalar field equation in flat space time is
\begin{eqnarray}
\partial_\mu \partial^\mu \psi +\lambda u^\alpha u^\beta \partial_\alpha \partial_\beta \psi +m^2\psi=0.
\end{eqnarray}

Further, the action of scalar particles based on Lorentz violationg scalar field theory is
\begin{eqnarray}
\mathbf{S}=\int d^4x\sqrt{-g}\frac{1}{2}[\partial_\mu \psi \partial^\mu\psi+\lambda(u^\alpha\partial_\alpha \psi)^2+m^2\psi^2].
\end{eqnarray}
The condition given in Eq. (2) holds true for the vector $u^\alpha$ in the above equation. Taking into account the electromagnetic potential, the action of scalar particles is 
\begin{eqnarray}
\mathbf{S}&=&\int d^4x\sqrt{-g}\frac{1}{2}\{(\partial_\mu +ieA_mu)\psi(\partial^\mu-ieg^{\mu \nu}A_\nu)\psi\cr && +\lambda u^\alpha(\partial_\alpha+)\psi u^\beta(\partial_\beta-ieA_\beta)\psi+m^2\psi^2\}.
\end{eqnarray}
The scalar field equation under Lorentz violation in curved space time can be written from the above equation as
\begin{eqnarray}
-\frac{1}{\sqrt{-g}}(\partial_\mu-ieA_\mu)[\sqrt{-g}(g^{\mu\nu}+\lambda u^\mu u^\nu)(\partial_\nu-ieA_\nu)\psi]+m^2\psi=0, 
\end{eqnarray}
where $A_\mu$ is the electromagnetic potential of the black hole and $e$ is the charge  of the particle.
Eq. (6) is the modified form of Klein-Gordon equation in curved space time induced by Lorentz violation theory. Equating the proportionality constant $\lambda$ to zero, we get the original Klein-Gordon equation coupled with an electromagnetic potential. In order to obtain the modified Hamilton-Jacobi equation in curved space time, we can write the wave function $\psi$ of Eq. (6) as
\begin{eqnarray}
\psi=\psi_0 \mathbf{e}^{\frac{i }{\hbar}S},
\end{eqnarray}
where $S(t,r,\theta,\phi)$ is the Hamilton principle function of scalar particles and $\hbar$ is the Planck's constant. Substituting Eq. (7) in Eq. (6) and neglecting the higher order terms of $\hbar$ in accordance with the semiclassical approximation, we get
\begin{eqnarray}
(g^{\mu\nu}+\lambda u^\mu u^\nu)(\partial_\mu S-eA_\mu)(\partial_\nu S-eA_\nu)+m^2=0.
\end{eqnarray}
Eq. (8) is the modified form of Hamilton-Jacobi equation of scalar field particles in curved space time induced by Lorentz violation.
This will be used to investigate the modified Hawking temperatures and change in Bekenstein-Hawking entropy in the later sections.

\section{Hawking temperature of Riemann space time for scalar particle}

The line element of the static Riemann space time, as given in Ref. \cite{ch1}, is
\begin{eqnarray}
ds^2&=&-a^2dt^2+b^2dx^2+c^2dy^2+d^2dz^2,
\end{eqnarray}
 where $a,b,c$ and $d$ are the functions of $(x,y,z)$ and the location of the event horizon of Eq. (9) is at $x=\xi$. The contravariant and covariant components of Riemann space time, as given in Ref. \cite{ch2}, can be written as
\begin{eqnarray}
g_{00}&=&-q^2(x-\xi)=-a^2 ,\cr  g^{11}&=&p^2(x,y,z)(x-\xi)=\frac{1}{b^2},\cr 
g^{22}&=&\theta,\cr
g^{33}&=&\phi,
\end{eqnarray}
where $q^2, p^2, \theta^2$ and $\phi$ are arbitrary functions which are not zero and nonsingular at the event horizon.
According to Ref. \cite{ch2}, the surface gravity and the Hawking temperature near the event horizon of Riemann space time are calculated as 
\begin{eqnarray}
\kappa=\lim_{g_{00} \to 0}\frac{1}{2}\sqrt{-\frac{g_{11}}{g_{00}}}\frac{\partial g_{00}}{\partial x}=\frac{1}{2}p(\xi)q(\xi)
\end{eqnarray} and 
\begin{eqnarray}
T_0=\frac{p(\xi)q(\xi)}{4\pi}
\end{eqnarray}
respectively.
To study the modified Hawking temperature, the ether-like field $u^\alpha$ are constructed as
\begin{eqnarray}
u^t&=&\frac{c_t}{\sqrt{-g_{tt}}}=\frac{c_t}{a}, \cr u^x&=&\frac{c_x}{\sqrt{g_{xx}}}=\frac{c_x}{b}, \cr u^y&=&\frac{c_y}{\sqrt{g_{yy}}}=\frac{c_y}{c}, \cr u^z&=&\frac{c_z}{\sqrt{g_{zz}}}=\frac{c_z}{d},
\end{eqnarray}
where $c_t,c_x, c_y$ and $c_z$ are arbitrary constants and $u^\alpha$ satisfies the following property
\begin{eqnarray}
u^\alpha u_\alpha=-c_t^2+c_x^2+c_y^2+c_z^2={\rm constant}.
\end{eqnarray}
Using Eqs. (9) and (13) in Eq. (8), we get
\begin{eqnarray}
&&-\frac{1}{a^2}(1+\lambda c_t^2)\left(\frac{\partial S}{\partial t}\right)^2+\frac{1}{b^2}(1+\lambda c_x^2)\left( \frac{\partial S}{\partial x}\right)^2+\frac{1}{c^2}(1+\lambda c_y^2)\left( \frac{\partial S}{\partial y}\right)^2\cr && +\frac{1}{d^2}(1+\lambda c_z^2)\left( \frac{\partial S}{\partial z}\right)^2+\left(\frac{2\lambda c_t c_x}{ab}\frac{\partial S}{\partial t}+\frac{2\lambda c_x c_y}{bc}\frac{\partial S}{\partial y}+\frac{2\lambda c_x c_z}{bd}\frac{\partial S}{\partial z}\right)\frac{\partial S}{\partial x}\cr &&+ \frac{2\lambda c_t c_x}{ad}\frac{\partial S}{\partial t}\frac{\partial S}{\partial z}+\frac{2\lambda c_t c_y}{ac}\frac{\partial S}{\partial t}\frac{\partial S}{\partial y} +\frac{2\lambda c_y c_z}{cd}\frac{\partial S}{\partial c}\frac{\partial S}{\partial d} + m^2=0.
\end{eqnarray}
To study the modified Hawking temperature of Riemann space time in the Lorentz violation theory, the action $S$ can be written as
\begin{eqnarray}
S=-\omega t+R(x)+W(y,z)+\epsilon,
\end{eqnarray}
where $\omega$ is the energy of the emitted scalar particle and $\epsilon$ is a complex constant. Using Eq. (16) in Eq. (15), we obtain a quadratic equation in $\frac{\partial R}{\partial x}$ as
\begin{eqnarray}
A_1\left(\frac{\partial R}{\partial x}\right)^2+B_1\frac{\partial R}{\partial x}+ C_1= 0,
\end{eqnarray}
where 

\begin{eqnarray}
A_1&=&p^2(x-\xi)(1+\lambda c_x^2), \cr
B_1&=&-\frac{2p\lambda c_t c_x \omega }{q} + 2\lambda c_x c_y \sqrt{\theta}\sqrt{p^2(x-\xi)}\frac{\partial W}{\partial y}+ 2\lambda c_x c_z \sqrt{p^2(x-\xi)}\sqrt{\phi}\frac{\partial W}{\partial z}, \cr
C_1&=& \theta \left(\frac{\partial W}{\partial y}\right)^2+\lambda \theta c_y^2\left(\frac{\partial W}{\partial y}\right)^2-\frac{2\lambda c_t c_y \omega \sqrt{\theta}}{\sqrt{q^2(x-\xi)}}\frac{\partial W}{\partial y}+2\lambda c_y c_z \sqrt{\theta}\sqrt{\phi}\frac{\partial W}{\partial y}\frac{\partial W}{\partial z} \cr && -\frac{1}{q^2(x-\xi)}\omega^2+\phi \left(\frac{\partial W}{\partial z}\right)^2+\frac{\lambda c_t^2}{q^2(x-\xi)}\omega^2-\frac{2\lambda c_t c_z \sqrt{\phi}}{\sqrt{q^2(x-\xi)}} \cr && +\lambda c_z^2 \phi \left(\frac{\partial W}{\partial z}\right)^2+m^2.
\end{eqnarray}
 From Eq. (17), we obtain the two roots as
\begin{eqnarray}
{\rm R}_\pm=\int \frac{\omega \left(\lambda c_t c_x  \pm \sqrt{1-\lambda c_t^2+\lambda c_x^2}\right)}{pq(x-\xi)(1+\lambda c_x^2)} dr,
\end{eqnarray}
where $R_+$ and $R_-$ correspond to scalar particles moving away from and  approaching the black hole respectively. Completing the above integral using Feynmann prescription, we get
\begin{eqnarray}
{\rm Im}R_\pm =\frac{ \pi\omega{\left(\lambda c_t c_x  \pm \sqrt{1-\lambda c_t^2 +\lambda c_x^2}\right)} }{p(\xi)q(\xi)(1+\lambda c_x^2)}.
\end{eqnarray}

The probabilities of ingoing and outgoing particles across the event horizon of the black hole are given by
\begin{eqnarray}
{\rm Prob(emmission)}=\exp (-2{\rm ImS})=\exp[-2({\rm ImR_+} + {\rm Im\epsilon})]
\end{eqnarray}
and
\begin{eqnarray}
{\rm Prob(absorption)}=\exp({\rm -2ImS}=\exp[-2({\rm ImR_-+Im\epsilon})].
\end{eqnarray}
Since the ingoing particle has a $100\%$ chance of entering the black hole in accordance with the semiclassical approximation, it indicates that ${\rm Im\epsilon=-Im R_-}$. The tunneling probability of the scalar particles at the event horizon $x=\xi$  of Riemann space time is
\begin{eqnarray}
\Gamma=\frac{\rm {Prob(emission)}}{\rm {Prob(absorption)}}=\exp\left( -\frac{ 4\pi\omega\sqrt{1-\lambda c_t^2 +\lambda c_x^2}}{p(\xi)q(\xi)(1+\lambda c_x^2 )}\right).
\end{eqnarray}
The modified Hawking temperature is given by
\begin{eqnarray}
T=\frac{p(\xi)q(\xi)}{4\pi \beta}=\frac{T_0}{\beta},
\end{eqnarray}
where
\begin{eqnarray}
\beta=\frac{\sqrt{1-\lambda c_t^2 +\lambda c_x^2}}{1+\lambda c_x^2}.
\end{eqnarray} 
Thus, we obtain the modified Hawking temperature of Riemann space time due to the Lorentz violation theory in curved space time. Here, $T_0$ is the original Hawking temperature of the Riemann space time. The modified Hawking temperature near the event horizon $x=\xi$ of Riemann space time increases or decreases if $0<\beta<1$ or $1<\beta<\infty$ respectively . If $\beta$ tends to 1, the actual Hawking temperature is recovered. It is observed that this modified Hawking temperature not only depends on the Lorentz violation parameter $\lambda$ but also on the ether-like vectors $u^\alpha$. 

\section{Tunneling of uncharged non-rotating BTZ black hole } 

The non-rotating uncharged BTZ black hole in spherical coordinates, as given by Ref. \cite{far}, is
\begin{eqnarray}
ds^2= Z(r)dt^2-Z^{-1}(r)dr^2-r^2d\phi^2,
\end{eqnarray}
where
\begin{eqnarray}
Z(r)=-M+\frac{r^2}{l^2}.
\end{eqnarray}
Eq. (26) has a singularity at $Z(r)=0$ and the value of black hole mass $M$ is
\begin{eqnarray}
M=\frac{r^2}{l^2}.
\end{eqnarray}
According to Ref. \cite{far}, the Hawking temperature of BTZ black hole is given as
\begin{eqnarray}
{\rm T_0'}=\frac{\sqrt{M}}{2\pi l}.
\end{eqnarray}
To study the modified Hawking temperature, the ether-like vectors $u^\alpha$ are constructed as
\begin{eqnarray}
u^t&=&\frac{c_t}{\sqrt{g_{tt}}}=\frac{c_t}{\sqrt{Z(r)}}, \cr u^r&=&\frac{c_r}{\sqrt{-g_{rr}}}=c_r\sqrt{Z(r)},\cr u^\phi&=&\frac{c_\phi}{\sqrt{-g_{\phi \phi}}}=\frac{c_\phi}{r},
\end{eqnarray}
where $c_t,c_r,$ and $c_\phi$ are arbitrary constants and $u^\alpha$ satisfies the following property
\begin{eqnarray}
u^\alpha u_\alpha=c_t^2-c_r^2-c_\phi^2={\rm constant}.
\end{eqnarray}
Using Eqs. (26) and (30) in Eq. (8), we obtain
\begin{eqnarray}
&&Z(r)(\lambda c_r^2-1)\left(\frac{\partial S}{\partial r}\right)^2+\frac{1}{Z(r)}(1+\lambda c_t^2)\left( \frac{\partial S}{\partial t}\right)^2+\frac{1}{r^2}(\lambda c_\phi^2-1)\left(\frac{\partial S}{\partial\phi}\right)^2+\cr && 2\lambda c_t c_r \frac{\partial S}{\partial t}\frac{\partial S}{\partial r}+2\lambda c_r c_\phi \frac{\sqrt{Z(r)}}{r}\frac{\partial S}{\partial \phi}\frac{\partial S}{\partial r}+2\lambda c_t c_\phi \frac{1}{\sqrt{Z(r)}r}\frac{\partial S}{\partial t}\frac{\partial S}{\partial \phi}+m^2=0.
\end{eqnarray}
Since Eq. (32) contains the variables $r$, $t$ and $\phi$, the action $S$ can be written as
\begin{equation}
S=-\omega t+R(r)+j\phi+\alpha,
\end{equation}
where $\omega$, $j$ and $\alpha$ are the energy of the emitted particle, angular momentum and complex constant respectively. Using Eq. (33) in Eq. (32), we obtain a quadratic equation in $\frac{\partial R}{\partial r}$ as 
\begin{eqnarray}
A_2\left(\frac{\partial R}{\partial r}\right)^2+B_2\frac{\partial R}{\partial r}+ C_2 = 0,
\end{eqnarray}
where
\begin{eqnarray}
A_2&=&Z(r){\lambda c_r^2-1}, \cr
B_2&=&-2\lambda c_t c_r \omega + 2\lambda c_r c_\phi \frac{\sqrt{Z(r)}}{r},\cr
C_2&=& \frac{1}{Z(r)}(1+\lambda c_t^2) \omega^2 + \frac{1}{r^2}(\lambda c_\phi^2-1)j^2-\frac{2\lambda c_t c_\phi \omega j}{\sqrt{Z(r)}r}+m^2.
\end{eqnarray}
The two roots of Eq. (34) are given by
\begin{eqnarray}
{\rm R}_\pm=\int \frac{\omega \left(\lambda c_t c_r \pm \sqrt{1+\lambda c_t^2-\lambda c_r^2}\right)}{Z(r)(\lambda c_r^2-1)} dr,
\end{eqnarray}
where $R_+$ and $R_-$ are the scalar particles moving away from and  approaching the black hole respectively. Solving the above integral, we have
\begin{eqnarray}
{\rm Im}R_\pm= \frac{\pi \omega l (\lambda c_t c_r\pm \sqrt{1+\lambda c_t^2-\lambda c_r^2)}}{2\sqrt{M}(\lambda c_r^2-1)}.
\end{eqnarray}

The tunneling probability of scalar particles across the black hole event horizon is
\begin{eqnarray}
\Gamma={\rm \frac{Prob(emmission)}{Prob(absorption)}}=\exp \left[ -\frac{2\pi \omega l \sqrt{1+\lambda c_t^2-\lambda c_r^2}}{\sqrt{M}(\lambda c_r^2-1)}\right].
\end{eqnarray}
The modified Hawking temperature of uncharged non-rotating BTZ black hole is calculated as
\begin{eqnarray}
{\rm T}=\frac{\sqrt{M}}{2\pi l \gamma}=\frac{T_0'}{\gamma},
\end{eqnarray}
where
\begin{eqnarray}
\gamma=\frac{\sqrt{1+\lambda c_t^2-\lambda c_r^2}}{\lambda c_r^2-1}.
\end{eqnarray}
Thus, we obtain the modified Hawking temperature due to the Lorentz violation theory in curved space time. Here, $T_0'$ is the original Hawking temperature of the uncharged non-rotating BTZ black hole. The modified Hawking temperature of non-rotating BTZ black hole increases or decreases if $0<\gamma<1$ or $1<\gamma<\infty$. If $\gamma$ tends to 1, the Lorentz violation is cancelled and the actual Hawking temperature is recovered. It is observed that this modified Hawking temperature not only depends on the Lorentz violation parameter $\lambda$ but also on the ether-like vectors $u^\alpha$.

\section{Tunneling of charged non-rotating BTZ black hole} 

The line element of a charged non-rotating BTZ black hole with mass $M$ and charge $Q$ is given in Refs. \cite{ban, mar, clem} as
\begin{eqnarray}
ds^2= Y(r)dt^2-\frac{1}{Y(r)}dr^2-r^2d\phi^2,
\end{eqnarray}
where
\begin{eqnarray}
Y(r)=\left(\frac{r^2}{l^2}-M-\frac{Q^2}{2} \ln \left(\frac{r}{l}\right)\right).
\end{eqnarray}
The electromagnetic potential of BTZ black hole is 
\begin{eqnarray}
A_\mu=(A_t,0,0),
\end{eqnarray} 
where $A_t=-Q\ln\left(\frac{r_+}{l}\right)$. The locations of the horizons are given by
\begin{eqnarray}
\frac{r_\pm^2}{l^2}-M-\frac{Q^2}{2}\ln\frac{r_\pm}{l}=0,
\end{eqnarray}
where $r_+$ and $r_-$ denote the event and Cauchy horizon respectively.
According to Ref. \cite{ejaz}, the Hawking temperature of the charged non-rotating BTZ black hole is given as
\begin{eqnarray}
T_h=\frac{1}{4\pi}\left(\frac{2r_+}{l^2}-\frac{Q^2}{2r_+}\right).
\end{eqnarray}
The heat capacity of the charged non-rotating BTZ black hole is calculated as
\begin{eqnarray}
C_h&=&T_h\left(\frac{\partial S}{\partial  T_h}\right)\cr &=&\frac{8\pi^2l^2T_h}{1+\frac{Q^2l^2}{4r_+}}.
\end{eqnarray}
By considering the limit of the outer horizon, $Y(r)$ can be written as 
\begin{eqnarray}
Y(r_+)=\left(\frac{2r_+}{l^2}-\frac{Q^2}{2r_+}\right)(r-r_+)
\end{eqnarray}
To study the modified Hawking temperature, the ether-like vectors $u^\alpha$ are constructed as
\begin{eqnarray}
u^t&=&\frac{c_a}{\sqrt{g_{tt}}}=\frac{c_a}{\sqrt{Y(r)}} , \cr u^r&=&\frac{c_b}{\sqrt{-g_{rr}}}=c_b \sqrt{Y(r)},\cr u^\phi&=&\frac{c_c}{\sqrt{-g_{\phi \phi}}}=\frac{c_c}{r},
\end{eqnarray}
where $c_a$, $c_b$ and $c_c$ are arbitrary constants and $u^\alpha$ satisfies the following property
\begin{eqnarray}
u^\alpha u_\alpha=c_a^2-c_b^2-c_c^2={\rm constant}.
\end{eqnarray}
Using Eqs. (41) and (48) in Eq. (8), we obtain
\small
\begin{eqnarray}
&&Y(r)(\lambda c_b^2-1)\left(\frac{\partial S}{\partial r}\right)^2+\frac{1}{Y(r)}(1+\lambda c_a^2)\left(eA_t- \frac{\partial S}{\partial t}\right)^2+\frac{1}{r^2}(\lambda c_c^2-1)\left(\frac{\partial S}{\partial\phi}\right)^2+m^2+ \cr && \left(2\lambda c_a c_b \frac{\partial S}{\partial t} +2\lambda c_b c_c \frac{\sqrt{Y(r)}}{r}\frac{\partial S}{\partial \phi}-2\lambda c_a c_b eA_t\right)\frac{\partial S}{\partial r}+2\lambda c_a c_c \frac{1}{\sqrt{Y(r)}r}\frac{\partial S}{\partial t}\frac{\partial S}{\partial \phi}=0.
\end{eqnarray}
\normalsize
To obtain the modified Hawking temperature and heat capacity, the action $S$ in Eq. (50) can be written as
\begin{equation}
S=-\omega t+R(r)+j\phi+\delta
\end{equation}
where $\omega$ and $j$ denote the energy and angular momentum of the emitted particle. $\delta$ is a complex constant. Using Eq. (51) in Eq. (50), a quadratic equation in $\frac{\partial R}{\partial r}$ is obtained as
\begin{eqnarray}
A_3\left(\frac{\partial R}{\partial r}\right)^2+B_3\frac{\partial R}{\partial r}+ C_3 = 0,
\end{eqnarray}
where
\begin{eqnarray}
A_3&=&Y(r)(\lambda c_b^2-1), \cr
B_3&=&-2\lambda c_a c_b \omega + \frac{2\lambda c_b c_c \sqrt{Y(r)}j}{r}-2\lambda c_a c_b eA_t,\cr
C_3&=& \frac{1}{Y(r)}(1+\lambda c_a^2) \omega^2 + \frac{1}{r^2}(\lambda c_c^2-1)j^2-\frac{2\lambda c_a c_c \omega j}{\sqrt{Y(r)}r}+m^2.
\end{eqnarray}
Solving for Eq. (52), the two roots obtained are given by
\begin{eqnarray}
{\rm R}_\pm=\int\frac{\lambda c_a c_b (\omega+eA_t) \pm \sqrt{(\omega+eA_t)^2(1+\lambda c_a^2-\lambda c_b^2)}}{Y(r)(\lambda c_b^2-1)}dr,
\end{eqnarray}
where $R_+$ and $R_-$ denote the scalar particles moving away from and approaching the black hole respectively. Solving the above integral, we obtain
\begin{eqnarray}
{\rm Im}R_\pm=\frac{\pi (\omega+eA_t) (\lambda c_a c_b\pm \sqrt{1+\lambda c_a^2-\lambda c_b^2})}{\left(\frac{2r_+}{l^2}-\frac{Q^2}{2r_+}\right)(\lambda c_b^2-1)},
\end{eqnarray}
The tunneling probability of the scalar particle across the black hole event horizon is
\begin{eqnarray}
\Gamma={\rm \frac{Prob(emmission)}{Prob(absorption)}}=\exp\left[-\frac{4\pi(\omega+eA_t)  \sqrt{1+\lambda c_a^2-\lambda c_b^2}}{\left(\frac{2r_+}{l^2}-\frac{Q^2}{2r_+}\right)(\lambda c_b^2-1)}\right].
\end{eqnarray}
The modified Hawking temperature is given by
\begin{eqnarray}
{\rm T}&=&\frac{1}{4\pi\zeta}\left(\frac{2r_+}{l^2}-\frac{Q^2}{2r_+}\right) \cr &=&\frac{T_h}{\zeta}.
\end{eqnarray}
where
\begin{eqnarray}
\zeta=\frac{ \sqrt{1+\lambda c_a^2-\lambda c_b^2}}{\lambda c_b^2-1}.
\end{eqnarray}
Thus, we obtain the modified Hawking temperature of charged non-rotating BTZ black hole due to the Lorentz violation theory in curved spacetime. Here, $T_h$ is the original Hawking temperature of the charged non-rotating BTZ black hole. If $0<\zeta<1$ or $1<\zeta<\infty$, the modified Hawking temperature near the event horizon of the charged non-rotating BTZ black hole increases or decreases  respectively. If $\zeta$ tends to 1, the original value of Hawking temperature is recovered.

The modified heat capacity across the event horizon of the charged non-rotating BTZ black hole is calculated as 
\begin{eqnarray}
C_M&=&\frac{\partial M}{\partial T}\cr &=&\frac{\zeta 8\pi^2 l^2 \left(\frac{r_+^2}{l^2}-\frac{Q^2}{2r_+}\right)}{2\pi \left(1+\frac{Q^2 l^2}{4r_+^2}\right)} \cr &=&\zeta C_h.
\end{eqnarray}
where $C_h$ denotes the original heat capacity of the charged non-rotating BTZ black hole.

Thus, we obtain the modified heat capacity  due to the Lorentz violation theory in curved space time. If $0<\zeta<1$ or $1<\zeta<\infty$, the modified heat capacity of charged non-rotating BTZ black hole decreases or increases respectively. If $\zeta$ tends to 1, the original heat capacity is recovered. 

It is observed that this modified Hawking temperature and heat capacity not only depends on the Lorentz violation parameter $\lambda$ but also on the ether-like vectors $u^\alpha$.

\section{ Tunneling of Schwarzschild-de Sitter black hole}

The line element of the Swarzschild-de Sitter black hole (SdS) in spherical polar coordinates is given by Ref. \cite{ati} as
\begin{eqnarray}
ds^2&=&-\frac{\Delta}{r^2} dt^2+\frac{r^2}{\Delta} dt^2+r^2(d\theta^2 + \sin^2\theta d\theta^2),
\end{eqnarray}
where
\begin{eqnarray}
\Delta=\left(r^2-2mr-\frac{r^4}{l^2}\right).
\end{eqnarray}
Eq. (61) has two real roots $r_h$ and $r_c$ for $0<\eta=\frac{M^2}{l^2}<\frac{1}{27}$. The horizons of the SdS black hole, namely the event horizon and the cosmological horizon, are respectively located at
\begin{eqnarray}
&&r_h=\frac{2m}{\sqrt{3\eta}}\cos\frac{\pi+\psi}{3}, \cr && r_c=\frac{2m}{\sqrt{3\eta}}\cos\frac{\pi-\psi}{3},
\end{eqnarray}
where
\begin{eqnarray}
\psi=\cos^{-1}(3\sqrt{3\eta}).
\end{eqnarray}
When $\eta$ takes the value $\frac{1}{27}$, the two horizons $r_c$ and $r_h$ coincide. For $\eta<\frac{1}{27}$, we expand $r_h$ as
\begin{eqnarray}
r_h=2m\left(1+\frac{4m^2}{l^2}+...\right).
\end{eqnarray}

To study the modified entropy of black hole, the ether-like vectors $u^\alpha$ are constructed as
\begin{eqnarray}
u^t&=&\frac{c_d}{\sqrt{-g_{tt}}}=\frac{ c_d r}{\sqrt{\Delta}}, \cr  u^r&=&\frac{c_e}{\sqrt{g_{rr}}}=\frac{c_e \sqrt{\Delta}}{r}, \cr
u^\theta&=&\frac{c_f}{\sqrt{g_{\theta \theta}}}=\frac{c_f}{r} ,\cr
u^\phi&=&\frac{c_g}{\sqrt{-g_{\phi \phi}}}=\frac{c_g}{r\sin\theta},
\end{eqnarray}
where $c_d, c_e, c_f$ and $c_g$ are arbitrary constants and $u^\alpha$ satisfies the following property
\begin{eqnarray}
u^\alpha u_\alpha=-c_d^2+c_e^2+c_f^2+c_g^2={\rm constant}.
\end{eqnarray}
Using Eqs. (60) and (65) in Eq. (8), we obtain
\begin{eqnarray}
&&-\frac{1}{\Delta}(1+\lambda c_t^2)\left(\frac{\partial S}{\partial t}\right)^2+\Delta(1+\frac{\lambda c_r^2}{r^2})\left(\frac{\partial S}{\partial r}\right)^2+\frac{1}{r^2}(1+\lambda c_\theta^2)\left(\frac{\partial S}{\partial \theta}\right)^2\cr &&+\frac{1}{r^2 \sin^2\theta}(1+\lambda c_\phi^2)\left(\frac{\partial S}{\partial \phi}\right)^2+
\left(2\lambda c_t c_r \frac{\partial S}{\partial t}+\frac{2\lambda c_r c_\theta}{\sqrt{\Delta}r}\frac{\partial S}{\partial \theta}+\frac{2\lambda c_r c_\phi}{\sqrt{\Delta} r \sin\theta}\frac{\partial S}{\partial \phi}\right)\frac{\partial S}{\partial r}\cr && 
+\frac{2\lambda c_\theta c_\phi}{r^2\sin\theta}\frac{\partial S}{\partial \theta}\frac{\partial S}{\partial \phi}+\frac{2\lambda c_t c_\theta\sqrt{\Delta}}{r}\frac{\partial S}{\partial t}\frac{\partial S}{\partial \theta}+\frac{2\lambda c_t c_\phi \sqrt{\Delta}}{r \sin\theta}\frac{\partial S}{\partial t}\frac{\partial S}{\partial \phi}+m^2=0.
\end{eqnarray}
To derive the modified entropy of SdS black hole, the action $S$ in Eq. (67) can be written as
\begin{equation}
S=-\omega t+W(r,\theta)+j\phi.
\end{equation}
where $W(r,\theta)$, $\omega$  and $j$ are the generalised momentum, energy of the emitted particle and angular momentum. Using Eq. (68) in Eq. (67), we obtain a quadratic equation in $\frac{\partial W}{\partial r}$ as
\begin{eqnarray}
A_4\left(\frac{\partial W}{\partial r}\right)^2+B_4\frac{\partial W}{\partial r}+ C_4 = 0,
\end{eqnarray}
where
\begin{eqnarray}
A_4&=&\frac{\Delta}{r^2}\left(1+\lambda c_r^2 \right),\cr
B_4&=&2\lambda c_r c_\theta \frac{\sqrt{\Delta}}{r^2}\left(\frac{\partial W}{\partial \theta}\right)-2\lambda c_t c_r \omega +2\lambda c_r c_\phi j \frac{\sqrt{\Delta}}{r^2\sin^2\theta},\cr
C_4&=& \left(\frac{1}{r^2}+\frac{\lambda  c_\theta^2}{r^2}\right)\left(\frac{\partial W}{\partial\theta}\right)^2+\left(\frac{2\lambda c_\theta c_\phi j}{r^2 \sin\theta}-\frac{2\lambda c_t c_\theta \omega}{\sqrt{\Delta}}\right)\left(\frac{\partial W}{\partial \theta}\right) \frac{\omega^2 \lambda c_t^2 r^2}{\Delta}-\frac{\omega^2 r^2}{\Delta} \cr && +\frac{j^2}{r^2\sin^2\theta}-\frac{2\lambda c_t c_\phi \omega j}{\sqrt{\Delta \sin\theta}}+\frac{\lambda c_\phi^2 j^2}{r^2\sin^2\theta}+m^2.
\end{eqnarray}
Solving for Eq. (69), two roots having physical meanings are given as
\begin{eqnarray}
{\rm W}_\pm =\int \frac{\lambda c_t c_r \omega \pm \sqrt{\omega^2(1+\lambda c_r^2-\lambda c_t^2)}}{\frac{\Delta}{r^2}(1+\lambda c_r^2)}dr,
\end{eqnarray}
where $W_+$ and $W_-$ correspond to the outgoing and ingoing particle respectively. Solving the above integral for the outgoing particle using Feynmann prescription, we get
\begin{eqnarray}
{\rm Im}W_+&=& \frac{ \pi r_h^2\omega\left(\lambda c_t c_r + \sqrt{1+\lambda c_r^2-\lambda c_t^2}\right)}{\Delta,_ r(r_h)(1+\lambda c_r^2)} \cr 
&=&\frac{\pi r_h^2\omega\left(\lambda c_t c_r +\sqrt{1+\lambda c_r^2-\lambda c_t^2}\right)}{\left(r_h-m-2\frac{r_h^3}{l^2}\right)(1+\lambda c_r^2)},
\end{eqnarray}
Using Eq. (64) in the above equation, we get
\begin{eqnarray}
{\rm Im}W_+=\frac{\pi4m^2\left(1+\frac{4m^2}{l^2}+...\right)^2\left(\lambda c_t c_r + \sqrt{1+\lambda c_r^2-\lambda c_t^2}\right)}{\left[2m\left(1+\frac{4m^2}{l^2}+...\right)-m-\frac{2}{l^2}\left\{2m\left(1+\frac{4m^2}{l^2}+...\right)\right\}^3\right](1+\lambda c_r^2)}\omega .
\end{eqnarray}
There is a change in the mass of the SdS black hole from $m$ to $m-\omega$ when a particle with energy $\omega$ tunnels out. Since SdS black hole is non-rotating, the angular velocity of the particle at the horizon is zero. Therefore, the angular momentum is zero. Considering the self-gravitational interaction, we can compute the imaginary part of the action from Eq. (73)  in the integral form as
\begin{eqnarray}
{\rm Im}W_+ =\pi\sigma\int\limits_{0}^\omega{\frac{4m^2\left(1+\frac{4m^2}{l^2}+...\right)^2}{2m\left(1+\frac{4m^2}{l^2}+...\right)-m-\frac{2}{l^2}\left\{2m\left(1+\frac{4m^2}{l^2}+...\right)\right\}^3}d\omega'},
\end{eqnarray}
where
\begin{eqnarray}
\sigma=\frac{\lambda c_t c_r  \pm \sqrt{1+\lambda c_r^2-\lambda c_t^2}}{1+\lambda c_r^2}.
\end{eqnarray}

Replacing $m$ by $m-\omega$ in Eq. (74), we get

\begin{eqnarray}
{\rm Im}W_+ =-\pi\sigma\int\limits_{m}^{m-\omega}{\frac{4(m-\omega')^2X^2}{2(m-\omega')X-(m-\omega')-\frac{2}{l^2}\left\{2(m-\omega')X\right\}^3}d(m-\omega)}
\end{eqnarray}
where
\begin{eqnarray}
X=1+\frac{4(m-\omega')^2}{l^2}+... .
\end{eqnarray}
Neglecting the terms $(m-\omega)^j$ for $j\geq5$ and solving the above integral, we get
\begin{eqnarray}
{\rm Im}W_+=-\frac{\pi\sigma}{2}\left[ 4(m-\omega)^2\left(1+\frac{8(m-\omega)^2}{l^2}\right)-4m^2\left(1+\frac{8m^2}{l^2}\right)\right].
\end{eqnarray}

The tunneling probability for the black hole is
\begin{eqnarray}
\Gamma\sim \exp(-2 {\rm Im}W_+)&=&\exp\left\{\pi\sigma \left[ 4(m-\omega)^2\left(1+\frac{8(m-\omega)^2}{l^2}\right)-4m^2\left(1+\frac{8m^2}{l^2}\right)\right]\right\}\cr &=&\exp\left\{\pi\sigma(r_f^2-r_i^2)\right\}\cr &=&\exp\left[\sigma(\Delta S_{BH})\right],
\end{eqnarray}

where $r_f=2(m-\omega)\left(1+\frac{4(m-\omega)^2}{l^2}\right)$ and $r_i=2m\left(1+\frac{4m^2}{l^2}\right)$ are the locations of the event horizons of the black hole before and after the emission of the particle and $\Delta S_{BH}=[S_{BH}(m-\omega)-S_{BH}(m)]$ denotes the change in Bekenstein-Hawking entropy of SdS black hole in the absence of Lorentz violation theory. The change in Bekenstein-Hawking entropy of SdS black hole in the Lorentz violation theory is calculated as 
\begin{eqnarray}
\sigma \Delta S_{BH}=\sigma[S_{BH}(m-\omega)-S_{BH}(m)].
\end{eqnarray}
 If $0<\sigma<1$ or $1<\sigma<\infty$, the modified entropy of the SdS black hole decreases or increases respectively. If $\sigma$ tends to 1, Lorentz violation is eliminated and the actual change in Bekenstein-Hawking entropy is recovered. It is observed that this modified Bekenstein-Hawking entropy not only depends on the Lorentz violation parameter $\lambda$ but also on the ether-like vectors $u^\alpha$. 

\section{Discussion and Conclusion} 

The Hawking temperatures of Riemann space time, charged and uncharged non-rotating BTZ black holes are modified in the Lorentz violation theory as $\frac{T_0}{\beta}$,$\frac{T_0'}{\gamma}$ and $\frac{T_h}{\zeta}$. The modified Hawking temperatures increase or decrease near the event horizon of black holes if $0<\beta, \gamma, \zeta<1$ or $1<\beta, \gamma, \zeta < \infty$. In addition to the modified Hawking temperature, the heat capacity of the charged non-rotating BTZ black hole is also modified as $\zeta C_h$. The modified heat capacity increases or decreases as $1<\zeta<\infty$ or $0<\zeta<1$.

Figure 1 shows the graph of the original Hawking temperature and modified Hawking temperature of a charged non-rotating BTZ black hole for the parameters $Q=0.3$, $l=1$, $c_a=0.9$, $c_b=1.2$ and $\lambda=1$. Here, it is observed that the original temperature ($T_0$) and modified Hawking temperature ($T$) are zero at $r_+= 0.15$. When $0.15<r_+<\infty$, the original Hawking temperature $(T_0)$ is greater than the modified Hawking temperature $(T)$ and both are found to be positive. When $0<r_+<0.15$, then $T_0<T$ and both are found to be negative for the above set of parameters.

Consequently, the original heat capacity ($C_h$) and the modified heat capacity ($C_M$) of the charged non-rotating BTZ black hole are also illustrated in Figure 2 for the above same set of parameters. It is also observed that the original and modified heat capacity are zero at $r_+ = 0.15$. If $0.15<r_+<\infty$, the heat capacities are positive and $C_M>C_h$. If $0<r_+<0.15$, the heat capacities are negative and $C_M<C_h$.

\begin{figure}
    \centering
    \includegraphics[width=0.8\textwidth]{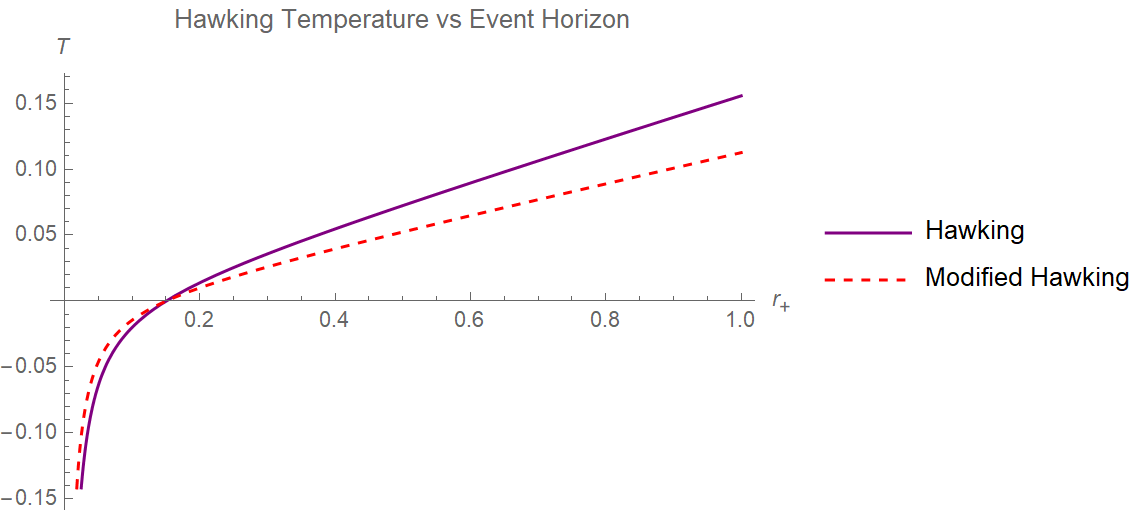}
    \caption{Plot of Hawking temperature and modified Hawking temperature as a function of the event horizon radius $r_+$ of a charged non-rotating BTZ black hole with $Q=0.3$, $l=1$ $c_a=0.9$, $c_b=1.2$ and $\lambda=1$. }
    \label{fig:my_label}
\end{figure}
\newpage
\begin{figure}
    \centering
    \includegraphics[width=0.8\textwidth]{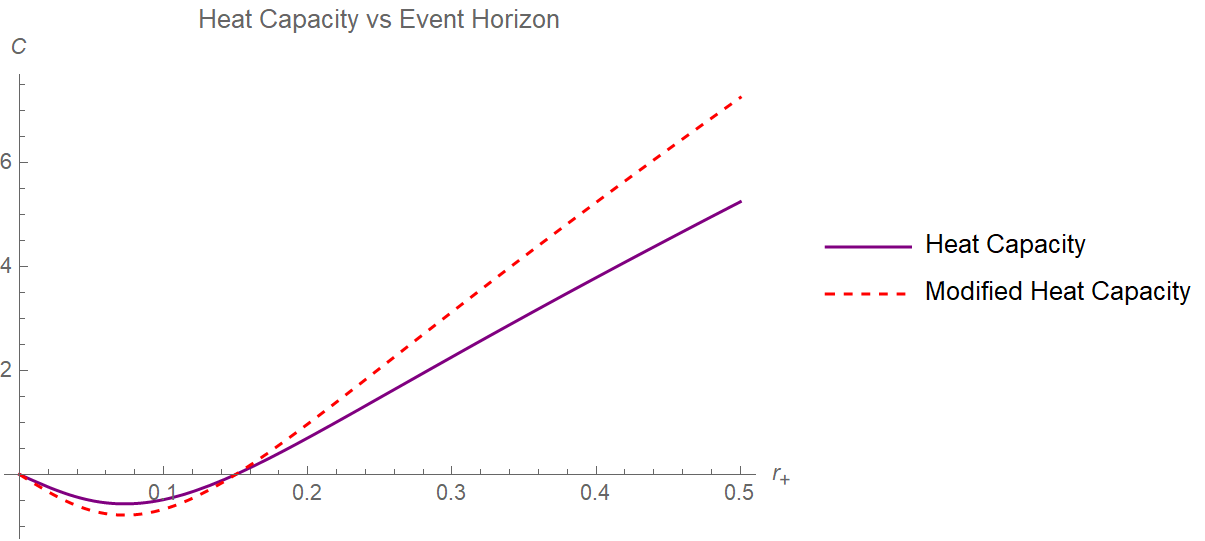}
    \caption{Plot of heat capacity and modified heat capacity as a function of the event horizon radius $r_+$ of a charged non-rotating BTZ black hole with $Q=0.3$, $l=1$, $c_a=0.9$, $c_b=1.2$ and $\lambda=1$.}
    \label{fig:my_label}
\end{figure}

The change in Bekenstein-Hawking entropy of SdS black hole is also studied using the Hamilton-Jacobi equation with Lorentz violation theory in curved space time. The modified Bekenstein-Hawking entropy increases or decreases depending on $\sigma\in(1,\infty)$ or $\sigma\in(0,1)$ respectively. 

In this paper, we investigate the tunneling of scalar particles across the event horizon of Riemann space time, BTZ black hole and SdS black hole by using the modified Hamilton-Jacobi equation with Lorentz violation theory in curved space time. The Hawking temperatures across the event horizon of Riemann space time, charged and uncharged non-rotating BTZ black hole are modified. The Bekenstein-Hawking entropy of the SdS black hole is also modified due to Lorentz violation theory. The modified Hawking temperatures and entropy depend on the values of $\beta$, $\gamma$, $\zeta$ and $\sigma$ which in turn depend on the values of Lorentz violation parameter $\lambda$ and  ether-like vectors $u^\alpha$.

For all the above cases, when the terms $\beta$, $\gamma$ and $\zeta$ tend to 1, the Lorentz violation theory is cancelled and the original Hawking temperatures of Riemann space time and BTZ black hole are recovered. Similarly,  the original Bekenstein-Hawking entropy of SdS black hole is also recovered when $\sigma$ tends to 1. It is worth mentioning that the modified Hawking temperatures and entropy depend not only on the value of Lorentz violation paramter $\lambda$ but also on the ether-like vectors $u^\alpha$. 

\section*{Acknowledgements}

First author acknowledges Council of Scientific and Industrial Research (CSIR), New Delhi for providing financial support.

\end{document}